\documentclass[manuscript]{acmart}

\usepackage{xspace}

\providecommand{\text}[1]{\mbox{#1}}
\providecommand{\mathbb}[1]{\mathbf{#1}}
\usepackage{subcaption} 

\begin{document}

\title[Affordable EEG, Actionable Insights]{Affordable EEG, Actionable Insights: An Open Dataset and Evaluation Framework for Epilepsy Patient Stratification}

\author{HM Shadman Tabib}
\affiliation{%
  \institution{Bangladesh University of Engineering and Technology (BUET)}
  \city{Dhaka}
  \country{Bangladesh}
}
\email{shadmantabib2002@gmail.com}

\author{Md. Hasnaen Adil}
\affiliation{%
  \institution{Bangladesh University of Engineering and Technology (BUET)}
  \city{Dhaka}
  \country{Bangladesh}
}

\author{Ayesha Rahman}
\affiliation{%
  \institution{Dhaka Medical College}
  \city{Dhaka}
  \country{Bangladesh}
}

\author{Ahmmad Nur Swapnil}
\affiliation{%
  \institution{Bangladesh University of Engineering and Technology (BUET)}
  \city{Dhaka}
  \country{Bangladesh}
}

\author{Maoyejatun Hasana}
\affiliation{%
  \institution{Bangladesh University of Engineering and Technology (BUET)}
  \city{Dhaka}
  \country{Bangladesh}
}

\author{Ahmed Hossain Chowdhury}
\affiliation{%
  \institution{Dhaka Medical College}
  \city{Dhaka}
  \country{Bangladesh}
}

\author{A. B. M. Alim Al Islam}
\affiliation{%
  \institution{Bangladesh University of Engineering and Technology (BUET)}
  \city{Dhaka}
  \country{Bangladesh}
}

\begin{abstract}
Access to clinical multi-channel EEG remains scarce in many regions worldwide. We present \textbf{NEUROSKY--EPI}, the first open dataset of single-channel, consumer-grade EEG for epilepsy, collected in a South Asian clinic together with rich contextual metadata. To probe its utility, we introduce \textit{EmbedCluster}, a patient-stratification pipeline that transfers representations from EEGNet trained on clinical recordings and augments them with contextual autoencoder embeddings, before applying unsupervised clustering to group patients by EEG patterns. Results indicate that low-cost, single-channel data can indeed support meaningful stratification. Beyond algorithmic performance, we foreground CHI concerns: deployability in resource-constrained settings, integration into everyday workflows, interpretability for non-specialists, and safeguards for privacy, inclusivity, and bias. By releasing the dataset and code, we aim to catalyze interdisciplinary work across HCI, health technology, and machine learning—charting a path toward affordable, actionable EEG that supports equitable epilepsy care
\end{abstract}

\maketitle

\section{Introduction}

Building on this vision, we now situate our work in the broader context of diagnosis and monitoring of epilepsy.
Epilepsy is a widespread neurological disorder that affects an estimated 50 million people worldwide \cite{WHO2019Epilepsy}. Continuous electroencephalography (EEG) remains the gold standard in clinical diagnosis and surveillance of seizures, but access to multichannel EEG systems is severely limited in many low and middle income regions due to high equipment costs, infrastructure demands, and the need for trained specialists \cite{Tatum2018EEGUtility}. These barriers often result in delayed diagnosis, irregular follow-up, and reduced access to effective treatment. Addressing this gap requires new approaches that balance clinical utility with accessibility. We therefore ask: \textit{What would epilepsy care look like if every rural clinic had a \$100 EEG stratifier?}

Recent work in human-computer interaction (HCI) has highlighted the growing potential of consumer-grade wearable EEG technologies to bring brain sensing to everyday contexts \cite{Casson2010WearableEEG, Patel2012WearableEEG}. Dry-electrode headsets integrated into familiar form factors, such as headphones, caps, or earbuds, reduce setup time, improve comfort, and enable data collection outside of traditional clinical environments. This shift opens opportunities for biosensing applications beyond the lab, including neurofeedback, interaction design, and health monitoring \cite{Zander2010TowardsPassive, Kosmyna2019AttentionAugmentation}. In particular, the integration of EEG into consumer devices supports human-centered goals of accessibility, usability, and inclusivity.

For epilepsy specifically, early feasibility studies suggest that low-cost, single-channel EEG headsets can capture physiologically meaningful markers of abnormal brain activity \cite{Casson2017SingleChannelEEG, NeuroSkyApplications}. Portable systems like the NeuroSky MindWave have been validated against clinical-grade EEG for tasks such as attention and meditation monitoring, demonstrating that consumer hardware can approximate traditional measurements \cite{NeuroSkyApplications}. More recent systematic reviews highlight the growing clinical interest in mobile EEG for seizure monitoring and management, suggesting that such approaches could complement or extend conventional hospital-based EEG systems \cite{Biondi2022MobileEEGReview}.

Despite these encouraging developments, significant gaps remain. First, most existing epilepsy datasets rely on multi-channel, hospital-grade EEG collected in controlled laboratory or clinical environments \cite{Goldberger2000PhysioNet}. Open datasets using consumer-grade, portable devices are nearly absent, creating barriers for researchers and designers interested in studying real-world, low-cost seizure monitoring. Second, many current machine learning approaches are data-intensive and assume large volumes of clean, multi-channel input. These assumptions limit generalizability and neglect the contextual factors—such as whether a patient is resting or alert—that can strongly influence EEG signals \cite{Craik2019DeepEEGReview, Roy2019DeepEEGSurvey, Lawhern2018EEGNet}.

To address these challenges, we introduce \textbf{NEUROSKY--EPI}, to our knowledge the first open-access epilepsy EEG dataset collected using a \emph{single-channel consumer-grade headset} in a South Asian low-resource context. We further propose \textit{EmbedCluster}, an unsupervised pipeline that combines EEGNet embeddings trained on clinical data with contextual autoencoder embeddings, then applies clustering to stratify patients by EEG patterns. Our contributions are as follows:

\begin{itemize}
    \item[\textbf{C.1}] \textbf{Open Dataset:} NEUROSKY--EPI, comprising 25 patients and over 2,000 labeled windows, with rich metadata including seizure types, medications, and demographics.
    \item[\textbf{C.2}] \textbf{Method:} An unsupervised clustering framework that extracts deep neural embeddings from EEGNet and complements them with contextual autoencoder embeddings.
    \item[\textbf{C.3}] \textbf{Evaluation:} Empirical demonstration of 62.50\% clustering accuracy with EEGNet embeddings and comparable performance (58.33--62.50\%) with contextual autoencoders, both exceeding chance (50\%).
    \item[\textbf{C.4}] \textbf{HCI Implications:} Reflection on usability, deployment, and ethics, highlighting how affordable EEG-based clustering could enable more inclusive epilepsy risk assessment in underserved communities.
\end{itemize}

Together, these contributions bridge machine learning rigor with HCI design concerns, laying the groundwork for future interdisciplinary work at the intersection of ubiquitous computing, health technology, and equitable care through intelligent patient stratification. To position these contributions within existing scholarship, we next review prior work on clinical EEG, consumer-grade alternatives, machine learning approaches, and open datasets.

\section{Related Work}

\textbf{Clinical EEG and Epilepsy Monitoring.} Electroencephalography remains the definitive diagnostic tool for epilepsy, with decades of research establishing its clinical utility \cite{Tatum2018EEGUtility, WHO2019Epilepsy}. Traditional multi-channel EEG systems, following the international 10–20 electrode placement system, capture complex spatiotemporal seizure patterns \cite{Goldberger2000PhysioNet}. However, these systems demand specialized training, controlled environments, and significant infrastructure \cite{Biondi2022MobileEEGReview}. While continuous EEG monitoring can improve detection rates by 50--80\% compared to routine short-term recordings \cite{Tatum2018EEGUtility}, access remains limited in low-resource settings. WHO estimates that up to 75\% of people with epilepsy in developing countries do not receive appropriate treatment, partly due to diagnostic limitations \cite{WHO2019Epilepsy}. This gap has motivated exploration of alternative monitoring approaches that balance diagnostic capability with accessibility and cost-effectiveness \cite{Casson2010WearableEEG, Patel2012WearableEEG}.  

\textbf{Wearable and Consumer EEG Technologies.} The emergence of dry-electrode EEG has enabled portable brain monitoring \cite{Casson2017SingleChannelEEG, Casson2010WearableEEG, Chi2010DryContact}. Unlike wet electrodes, dry sensors can be embedded into everyday objects (headphones, caps, earbuds), reducing setup time and improving comfort \cite{Fiedler2015NovelEEG, Wang2016DryElectrode}. Ambulatory systems such as smartphone-compatible ear electrodes \cite{Bleichner2016ConcealedEEG} achieve comparable signal quality to traditional scalp recordings. Consumer devices like the NeuroSky MindWave, Emotiv EPOC, and Muse headband have further broadened access \cite{Hairston2014UsabilityWearable}. Advanced systems now include wireless streaming \cite{Lin2010WirelessEEG}, real-time processing \cite{Mullen2015RealTimeEEG}, and flexible electronics \cite{Rogers2010FlexibleWireless, Norton2015SoftWireless}.  

HCI researchers have investigated EEG in ubiquitous computing environments \cite{Zander2010TowardsPassive, Kosmyna2019AttentionAugmentation}. Studies suggest that consumer EEG devices can often be operated with minimal training, enabling self-administered health monitoring \cite{Patel2012WearableEEG, Xu2017ReviewPortable}. Yet limitations remain: higher noise, lower spatial resolution, and variable reliability compared to clinical systems \cite{Raduntz2017EEGArtifact}. Still, validation work shows these devices can capture meaningful physiological markers (attention, meditation, basic neurological assessment) \cite{NeuroSkyApplications, Aspinall2006PortableEEG}. In epilepsy, mobile EEG offers shorter setup times, more naturalistic data collection, and earlier seizure detection \cite{Biondi2022MobileEEGReview, Ramgopal2014SeizureDetection}. These advances highlight feasibility but underscore the need for datasets beyond controlled clinical environments.  

\textbf{Machine Learning for Seizure Detection.} Automated seizure detection has progressed from hand-crafted features (spectral power \cite{Liang2010CombinationFeature}, entropy \cite{Kannathal2005EntropiesEEG}, wavelets \cite{Adeli2003WaveletEEG, Faust2015WaveletBased}, statistical moments \cite{Subasi2005ClassificationEMG}) with SVMs or ANNs \cite{Nicolaou2012DetectionSeizure, Gandhi2011ExpertModel} to deep learning. Compact CNNs such as EEGNet \cite{Lawhern2018EEGNet} learn directly from raw EEG, while RNNs \cite{Tsiouris2018LongShort}, attention models, and graph neural networks \cite{Covert2019TemporalGraphConvolutional} capture temporal and spatial dynamics. Ensemble methods \cite{Akyol2020AutomaticDetection} and 3D CNNs \cite{Wei2018AutomaticSeizure} have further advanced performance.  

However, most models assume multi-channel clinical EEG and are evaluated in controlled environments \cite{Roy2019DeepEEGSurvey, Shoeb2010PatientSpecific}. Domain adaptation remains a challenge: models trained on one population or device often degrade on another \cite{Craik2019DeepEEGReview, Andrzejak2001IndizesNonlinearDeterminism}. Transfer learning shows promise \cite{Boonyakitanont2020ReviewDeep, Rasheed2020MachineLearning}, yet few studies explore adapting from clinical EEG to consumer single-channel EEG. This gap motivates approaches like ours, which explicitly test such transfer.  

\textbf{Context-Aware and Personalized EEG.} EEG signals vary with arousal, cognitive state, time of day, and environment \cite{Williamson2012SeizureDetection, Temko2011EEGBased}. Patient-specific models often outperform population-based ones but require calibration data \cite{Shoeb2010PatientSpecific}. Mobile health systems increasingly integrate context (e.g., smartphone sensors) to personalize analysis, though EEG-specific applications remain sparse. Preprocessing methods such as signal transforms also help adapt algorithms to heterogeneous data \cite{San2019EpilepticSeizure}. These insights highlight the importance of contextual metadata—a key feature of our dataset.  

\textbf{Open Datasets and Ethics.} Public EEG datasets such as CHB-MIT \cite{Goldberger2000PhysioNet} have enabled progress but are predominantly clinical, multi-channel, and collected in hospital settings \cite{Roy2019DeepEEGSurvey}. Few datasets capture diversity of real-world usage, especially in low-resource contexts. This limits the development of inclusive, accessible monitoring systems.  

EEG data also raise ethical concerns. Privacy risks are significant: brain signals may reveal sensitive information about cognition or health. Algorithmic bias is another risk, as EEG patterns vary across demographics and models trained on homogeneous populations may underperform elsewhere \cite{Rajkomar2018EnsureFairnessML}. By including metadata on demographics, seizure types, and socioeconomic status, our dataset aims to support fairness analyses alongside technical work.

\section{System Overview and Data Collection}

We designed a \textbf{portable EEG monitoring system} centered on the NeuroSky MindWave Mobile 2 headset—an inexpensive, single-channel device worn on the forehead (Fp1) with a reference clip on the ear. The headset records frontal brain signals and internally computes band-power features (including proprietary ``Attention'' and ``Meditation'' metrics) in real time. While a single electrode cannot capture the full spatiotemporal complexity of seizures, the trade-off in \textit{usability and affordability} is substantial: patients can self-wear the device without technical assistance, making it well-suited for low-resource contexts where clinical EEG is often unavailable. These gaps underscore the need for accessible datasets and pipelines that both respect ethical concerns and operate with low-cost hardware. Our system overview describes how we sought to address exactly this challenge.

\textbf{Participant Cohort.} We collected data from 25 epilepsy patients (13 female, 12 male) at an anonymized hospital in South Asia. This cohort represents a diverse patient population in a low- and middle-income setting, with varying seizure types and socio-demographic backgrounds. Each participant provided informed consent (or assent with guardian consent for minors) under an Institutional Review Board-approved protocol. To capture different brain states, we recorded EEG in two controlled conditions: a \textit{resting phase} (eyes closed, relaxed, 1 minute) and an \textit{awake phase} (eyes open, engaged in light cognitive activity, 1 minute). The signals were segmented into $\sim$1-second windows, yielding a total of \textbf{2,032 labeled EEG windows} across 25 patients. Annotations reflect whether patients reported a \textbf{recent change in seizure frequency} (Yes = 15 patients, No = 10 patients). In addition, we collected de-identified \textbf{metadata} including demographics (age, gender, socioeconomic indicators, education), clinical variables (seizure types, medications, comorbidities, treatment duration), and self-reported measures (treatment satisfaction, adherence). All metadata fields are categorical or binned to prevent re-identification, supporting both technical evaluation and fairness analysis.

\begin{figure}[ht]
  \centering
  \begin{subfigure}[t]{0.8\linewidth}
    \centering
    \includegraphics[width=\linewidth]{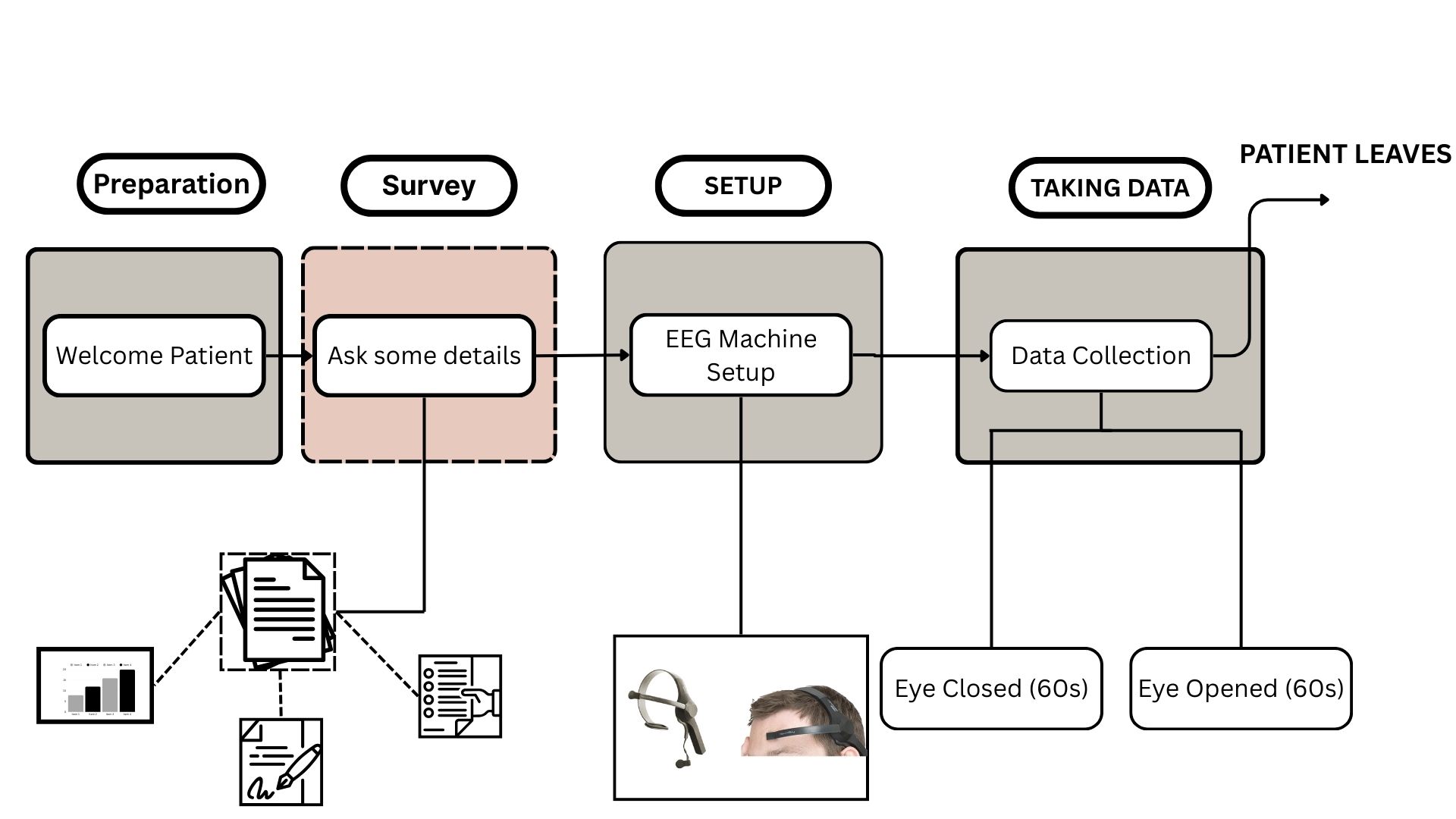}
    \subcaption{Study flow and protocol.}
    \label{fig:pt-flow}
  \end{subfigure}
  \hfill
  \begin{subfigure}[t]{0.5\linewidth}
    \centering
    \includegraphics[width=\linewidth]{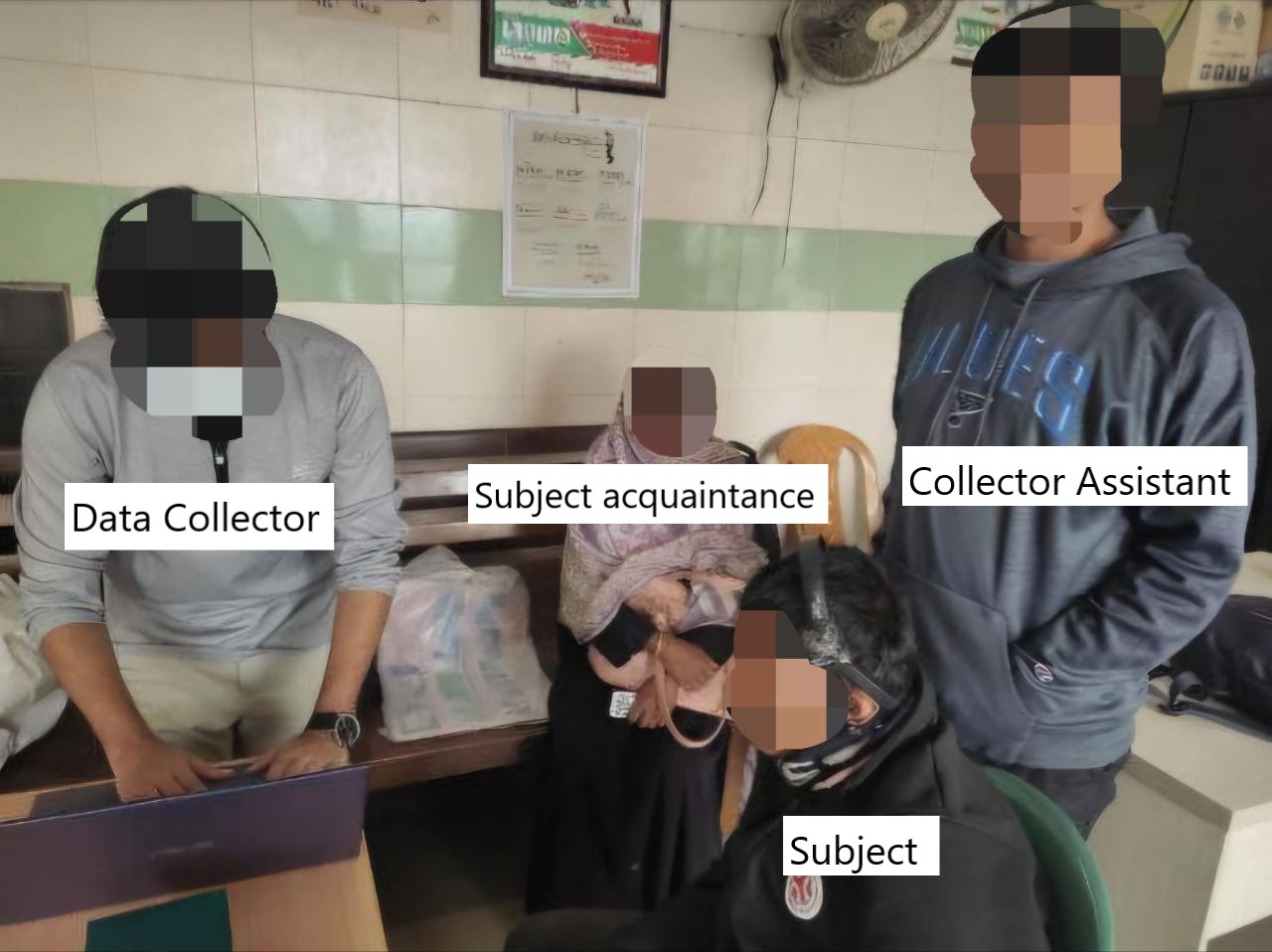}
    \subcaption{In-situ data collection (faces blurred).}
    \label{fig:taking-data}
  \end{subfigure}
  \caption{NEUROSKY--EPI data collection. \textbf{(a)} Process overview: after consent and survey, a single-channel consumer EEG is set up. Each participant completes two controlled conditions—\emph{resting/eyes-closed} (60s) and \emph{awake/eyes-open} (60s). Sessions are segmented into 1-second windows and paired with clinical metadata. \textbf{(b)} Photograph of a data collection session (faces blurred).}
  \label{fig:data-collection}
\end{figure}

\textbf{Training Data (CHB-MIT).} To build a seizure detection model for our single-channel setup, we leveraged the CHB-MIT Scalp EEG Database, which contains multi-channel recordings from 24 pediatric patients with annotated seizures. To approximate our consumer device, we extracted the Fp1 channel (or nearest frontal electrode) and segmented signals into 8-second windows, yielding 12,009 labeled segments. Each window was labeled as seizure or non-seizure per CHB-MIT annotations. These segments provided the supervised training data for our model. Crucially, \textbf{no NEUROSKY--EPI data was used for training}: it was held out entirely for cross-domain evaluation, ensuring results reflect generalization to a new device and population.

\begin{table}[ht]
\centering
\caption{Comparison of datasets used in this study. NEUROSKY--EPI complements clinical benchmarks like CHB-MIT by providing consumer-grade, single-channel EEG with contextual metadata.}
\begin{tabular}{p{3.2cm} p{3.5cm} p{3.5cm}}
\toprule
\textbf{Aspect} & \textbf{CHB-MIT} & \textbf{NEUROSKY--EPI} \\
\midrule
Population & 24 pediatric patients & 25 patients with epilepsy (South Asian hospital) \\
Setting & Hospital, clinical EEG & Hospital, consumer EEG in low-resource context \\
Hardware & Multi-channel scalp EEG (10--20 system) & Single-channel consumer headset (Fp1) \\
Duration & Long-term (hours per patient) & Short sessions (2 minutes per patient) \\
Data Volume & 12,009 windows (8s each) & 2,032 windows (1s each) \\
Annotations & Seizure vs. non-seizure & Seizure frequency change (Yes = 15, No = 10) \\
\bottomrule
\end{tabular}
\label{tab:dataset-comparison}
\end{table}

\textbf{System Workflow.} In practice, patients wear the MindWave headset, generating a stream of single-channel EEG. The headset converts signals into 10-dimensional band-power vectors every second (Delta through Gamma bands, plus attention and meditation metrics). These features can be transmitted to a smartphone or computer. Our pipeline then processes them—optionally integrating contextual features—to group patients by EEG patterns, with detected events logged as CSV files. This workflow illustrates how an affordable, portable device could fit into everyday hospital or community health contexts. The technical details of the embedding and clustering pipeline are described in the next section. 

\begin{figure}[ht]
\centering
\includegraphics[width=\textwidth]{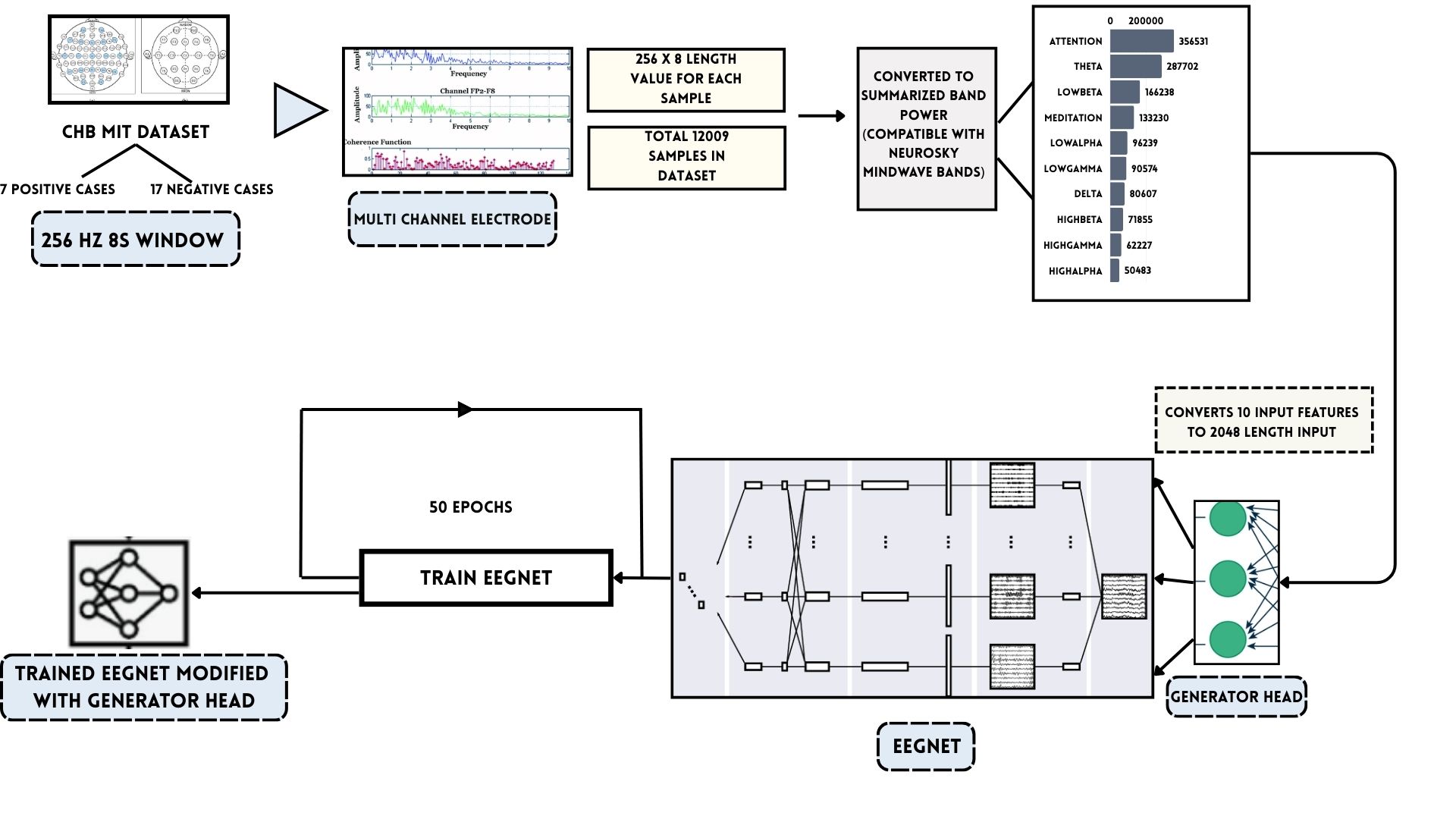}
\caption{Overview of our transfer learning approach (Stage~1 training). We train a seizure detection model on CHB-MIT, restricted to a single frontal channel to mirror the consumer headset. Raw 8-second EEG segments are converted into 10-dimensional band-power features to match the NeuroSky output. A \textbf{generator network} upsamples these features into a synthetic EEG sequence, which is processed by \textbf{EEGNet}, a compact CNN. EEGNet learns to classify seizure vs.\ non-seizure segments. This stage yields a calibrated model whose embeddings transfer to our consumer EEG domain.}
\label{fig:overview}
\end{figure}

\section{Methods}
Our technical approach centers on leveraging deep neural embeddings for \textbf{unsupervised patient stratification} with single-channel EEG. We call our pipeline \textbf{EmbedCluster}. Conceptually, the pipeline unfolds in two complementary stages: (1) \textbf{EEGNet embedding extraction}, where we pretrain a compact convolutional model on a clinical benchmark dataset (CHB-MIT) and then repurpose its internal feature representations for our consumer-grade EEG; and (2) \textbf{contextual autoencoder embeddings}, trained directly on NEUROSKY--EPI to emphasize dataset-specific and context-aware representations. These embeddings are then clustered with unsupervised algorithms (K-means, Agglomerative, Gaussian Mixture Models, Spectral clustering) to group patients into strata that reflect meaningful clinical differences. 

Our design rationale follows from two imperatives. First, \textit{transfer learning} allows us to bring the knowledge encoded in clinical EEG datasets into low-resource environments where annotated data are scarce. Second, \textit{contextual autoencoding} provides a bottom-up alternative that directly reflects the idiosyncrasies of our dataset without assuming clinical data availability. Together, these approaches operationalize Contributions \textbf{C.2--C.3} from the Introduction, providing a balanced exploration of knowledge transfer and self-contained modeling. Figures~\ref{fig:pipeline-nocontext} and \ref{fig:pipeline-context} provide schematic overviews of the two pipelines.

\subsection{EEGNet Embedding Extraction Pipeline}
The NeuroSky MindWave device provides per-window estimates of 10 frequency band features: \textit{Delta}, \textit{Theta}, \textit{Low-Alpha}, \textit{High-Alpha}, \textit{Low-Beta}, \textit{High-Beta}, \textit{Low-Gamma}, \textit{Mid-Gamma}, and two proprietary indices termed ``Attention'' and ``Meditation.'' These features represent the most basic frequency-domain summaries and are far less granular than raw multi-channel clinical EEG. Nevertheless, they are widely used in consumer-grade systems due to their real-time interpretability and low computational footprint. 

Formally, for each 1-second EEG window $x$, we obtain a feature vector $b(x) \in \mathbb{R}^{10}$. While these low-dimensional vectors are interpretable, they are not directly compatible with neural architectures like EEGNet, which expect high-resolution temporal sequences. To bridge this gap, we introduce a \textbf{generator head} network $g(\cdot)$ that expands the band features into a pseudo-EEG sequence. Concretely, $g: \mathbb{R}^{10} \to \mathbb{R}^{C \times T}$ where $C=1$ channel and $T=2048$ time-points (corresponding to an 8-second, 256~Hz signal). The generator is implemented as a lightweight multilayer perceptron that projects $b(x)$ into a 2048-length vector, which is reshaped into a $(1 \times 2048)$ pseudo-waveform. This choice reflects an HCI-oriented principle: \textit{minimal computational burden}, so that the method could plausibly run on smartphones or low-power embedded devices in deployment.  

We then feed this pseudo-waveform into EEGNet, a compact CNN designed for EEG decoding \cite{Lawhern2018EEGNet}. EEGNet uses temporal convolutions to capture frequency-specific patterns, depthwise convolutions to simulate spatial filters across channels (degenerating to $C=1$ in our case), and separable convolutions for efficient feature mixing. We adopt the EEGNet-8-2 variant, which strikes a balance between capacity and efficiency. \textbf{Importantly, rather than using EEGNet’s softmax outputs, we extract embeddings from the penultimate layer}. These embeddings (16 filters $\times$ time dimension) capture the representational richness of seizure-related neural dynamics without committing to specific clinical labels. This design mirrors prior work in representation learning, where penultimate embeddings often generalize better than task-specific outputs.

\subsection{Training EEGNet on CHB-MIT for Embedding Extraction}
Stage~1 training (Figure~\ref{fig:overview}) uses the CHB-MIT dataset as a source of clinically validated seizure vs.\ non-seizure labels. Each 8-second segment $x(t)$ from the Fp1 electrode (or nearest frontal proxy) is sampled at 256~Hz. To harmonize this high-resolution input with NeuroSky’s 10-dimensional feature format, we first apply \textbf{band-power extraction}. 

\paragraph{Band-power extraction.} Following the NeuroSky device’s design, we compute spectral power across ten canonical frequency bands:
\[
\mathcal{B} = \{\delta, \theta, \alpha_L, \alpha_H, \beta_L, \beta_H, \gamma_L, \gamma_H, \text{attention}, \text{meditation}\}.
\]
For each $b \in \mathcal{B}$, the band-power is estimated as:
\[
P_b = \frac{1}{T}\int_0^T \Big|\mathcal{F}\{x(t)\}_b\Big|^2 dt,
\]
with $\mathcal{F}\{x(t)\}_b$ denoting the Fourier transform restricted to band $b$, and $T=8$ seconds. The resulting vector
\[
\mathbf{z} = [P_{\delta}, P_{\theta}, P_{\alpha_L}, P_{\alpha_H}, P_{\beta_L}, P_{\beta_H}, P_{\gamma_L}, P_{\gamma_H}, P_{\text{att}}, P_{\text{med}}]
\]
replicates the NeuroSky’s output. The ``attention'' and ``meditation'' indices were approximated using empirically validated ratios of $\alpha$ and $\beta$ rhythms \cite{NeuroSkyApplications}.

\paragraph{Pipeline overview.} Summarizing:
\[
x(t)\ \xrightarrow{\ \text{band-power}\ }\ \mathbf{z}\in\mathbb{R}^{10}\ \xrightarrow{\ g(\cdot)\ }\ \tilde{x}\in\mathbb{R}^{2048}\ \xrightarrow{\ \text{EEGNet}\ }\ \hat{y}\in[0,1].
\]

\paragraph{Training details.} The pseudo-EEG signals $\tilde{x}$ were passed through EEGNet, with binary cross-entropy loss:
\[
\mathcal{L} = -\frac{1}{N}\sum_{i=1}^N \Big[y_i \log \hat{y}_i + (1-y_i)\log (1-\hat{y}_i)\Big].
\]
We trained for 50 epochs using Adam \cite{Kingma2014Adam} with learning rate $10^{-3}$ and batch size 64, applying early stopping. On CHB-MIT validation data (20\% split), the model achieved $\sim$75\% accuracy using only one frontal channel. This performance is lower than multi-channel baselines, but adequate for extracting embeddings transferable to consumer EEG. In alignment with reproducible ML practice, we fixed random seeds and logged hyperparameters for all runs.

\subsection{Patient-Level Embedding Extraction and Clustering}
\begin{figure}[ht]
\centering
\includegraphics[width=\textwidth]{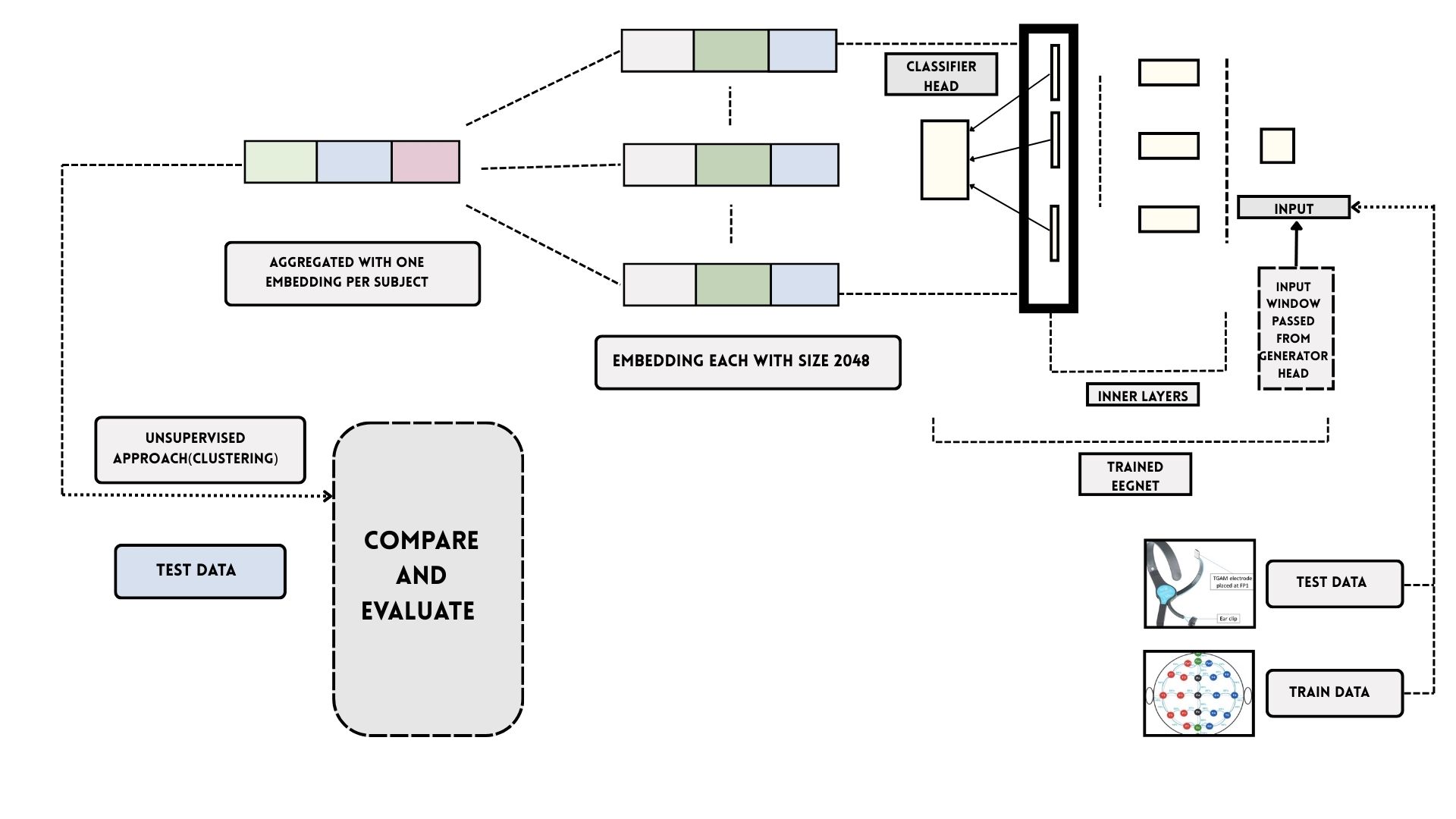}
\caption{Stage 2: EEGNet embedding extraction for unsupervised stratification. EEGNet embeddings are aggregated at the patient level (mean + std) and clustered. Clusters align with seizure frequency change annotations (Yes/No), achieving 62.50\% accuracy across algorithms.}
\label{fig:pipeline-nocontext}
\end{figure}

In Stage~2 (Figure~\ref{fig:pipeline-nocontext}), we apply the trained EEGNet to NEUROSKY--EPI. Each 1-second window is mapped via $f(x) = \text{EEGNet}(g(b(x)))$. From each forward pass, we collect the penultimate-layer embeddings (16$\times$1$\times$64 feature maps).

\paragraph{Patient-level aggregation.} For each patient, all embeddings are flattened (1024 dimensions per window). We then compute the mean and standard deviation across their windows, producing a 2048-dimensional patient-level embedding. This design captures both central tendency and variability: two properties often emphasized in clinical neurophysiology, where consistency and fluctuation both matter diagnostically.

\paragraph{Unsupervised clustering.} We cluster patients using four standard methods: K-means, Agglomerative, GMM, and Spectral clustering. Clustering quality is evaluated against our dataset’s annotation of \textbf{seizure frequency change} (Yes/No). The EEGNet embeddings consistently achieve \textbf{62.50\% accuracy} across all methods, substantially above chance (50\%). Importantly, cluster sizes remain balanced (12–13 per group), suggesting that embeddings capture clinically meaningful heterogeneity rather than overfitting to noise or outliers.

\subsection{Contextual Autoencoder Embedding Approach}
As a complement to transfer learning, we trained a contextual autoencoder directly on NEUROSKY--EPI \cite{Wen2018AECDNN, Li2015DenoisingAE, Zhang2020ExpressionEEG}. This choice reflects an HCI-inspired concern: what if external clinical datasets are unavailable, or cannot be legally or ethically transferred across borders? A purely local model ensures that the pipeline can still operate with community-collected data.

The autoencoder is a 3-layer feed-forward network with a code dimension of 4. To enrich its representational capacity, we appended 4 context summaries (mean band powers per recording condition: rest vs.\ active), yielding a 16-dimensional embedding $c(x)$. Training was unsupervised on all available windows, minimizing mean squared reconstruction error.

\paragraph{Patient-level clustering.} Embeddings $c(x)$ were aggregated per patient (mean + std), producing 32-dimensional patient representations. Clustering followed the same protocol as above. Results showed more variability: K-means and Spectral reached \textbf{58.33\%}, Agglomerative 58.33\%, while GMM peaked at \textbf{62.50\%}. Notably, some algorithms (e.g., Agglomerative) produced highly imbalanced clusters (21 vs.\ 4 patients), highlighting that autoencoder embeddings may emphasize dataset-specific idiosyncrasies rather than broadly generalizable features.  

\paragraph{Deployment relevance.} While accuracy is slightly lower on average, the autoencoder has a critical practical advantage: it does not depend on external clinical training data, making it better suited for self-contained deployments in under-resourced contexts. By contrast, the EEGNet approach leverages clinically grounded priors, yielding more structured embeddings. Figure~\ref{fig:pipeline-context} illustrates this comparison.

\begin{figure}[ht]
\centering
\includegraphics[width=0.9\textwidth]{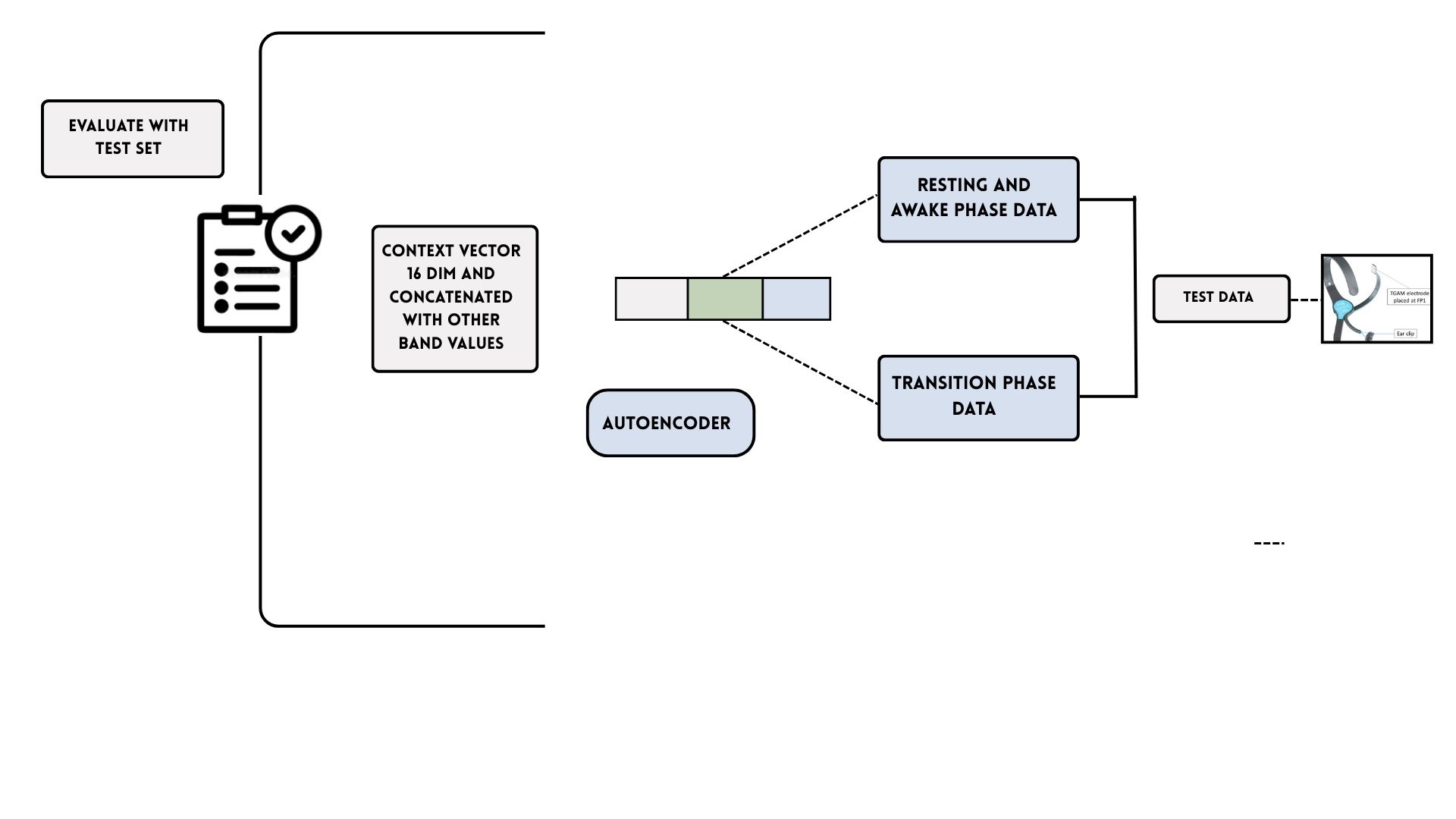}
\caption{Alternative approach: Autoencoder-based stratification. Contextual autoencoder embeddings were aggregated per patient and clustered. Accuracy ranged from 58.33\% to 62.50\%, with GMM best matching seizure frequency change annotations. This self-contained method avoids reliance on external data, highlighting its deployability in local contexts.}
\label{fig:pipeline-context}
\end{figure}

With both EEGNet-based and autoencoder-based pipelines in place, we next evaluate how well these representations support patient stratification on our dataset.

\section{Findings}
We evaluated our unsupervised clustering framework on the \textbf{NEUROSKY--EPI} dataset using two embedding strategies: (1) EEGNet embeddings extracted from a model pretrained on the clinical CHB-MIT dataset, and (2) contextual autoencoder embeddings trained directly on our consumer-grade dataset. In each case, we report clustering accuracy across multiple algorithms and interpret the results with respect to our broader design goals: enabling affordable patient stratification using single-channel EEG and examining the human-centered implications of such a system.

\subsection{EEGNet Embedding Clustering Performance}
Using EEGNet embeddings pretrained on CHB-MIT seizure vs.\ non-seizure data, we observed consistent and robust clustering outcomes across all tested algorithms. The EEGNet-based approach achieved \textbf{62.50\% clustering accuracy} across K-means, Agglomerative, GMM, and Spectral clustering algorithms, substantially above chance level (50\%). This robustness highlights that clinically trained models can transfer knowledge to a consumer EEG setting, capturing discriminative features even when applied to noisier, single-channel input.

Cluster distributions were notably well-balanced: for example, K-means produced clusters of 13 and 12 patients, while other algorithms yielded nearly identical splits (12–13 patients per cluster). Such balance suggests that the embeddings are not dominated by spurious device artifacts or random noise, but instead reflect meaningful patient-level distinctions. From an HCI perspective, this balance is critical: skewed or unstable clustering would be difficult to interpret and could undermine trust in a clinical workflow. Balanced clusters increase the likelihood that the tool could be integrated into everyday practice without raising immediate usability or fairness concerns.

\subsection{Autoencoder Embedding Clustering Performance}
The contextual autoencoder approach, trained exclusively on NEUROSKY--EPI data, displayed more variability across algorithms. K-means and Spectral clustering each achieved \textbf{58.33\% accuracy}, while GMM reached the highest accuracy at \textbf{62.50\%}, matching the EEGNet approach. Agglomerative clustering performed at a similar level (58.33\%), though with notable imbalance in patient groupings.

Cluster balance differed more dramatically than in the EEGNet setting. For example, Agglomerative clustering produced clusters of 21 and 4 patients, while K-means yielded a more even split of 15 and 10. This pattern suggests that the autoencoder embeddings may be more sensitive to dataset-specific noise or individual patient variation, capturing features that are distinctive but less generalizable. From a human-centered design standpoint, such imbalance matters: algorithms that disproportionately group most patients together may reduce interpretability for healthcare providers and risk overlooking clinically important subgroups.

\subsection{Summary of Results}
For clarity, Table~\ref{tab:results} summarizes the clustering accuracy achieved by each method across the four algorithms.

\begin{table}[ht]
\centering
\caption{Clustering accuracy (\%) for patient stratification on the \textbf{NEUROSKY--EPI dataset}. 
EEGNet embeddings were extracted by a model \emph{pretrained on CHB-MIT} and then applied to our dataset, 
while autoencoder embeddings were trained directly on NEUROSKY--EPI. Chance level = 50\%.}
\label{tab:results}
\begin{tabular}{lcccc}
\toprule
\textbf{Embedding Source} & \textbf{K-means} & \textbf{Agglomerative} & \textbf{GMM} & \textbf{Spectral} \\
\midrule
EEGNet embeddings (CHB-MIT $\rightarrow$ NEUROSKY--EPI) & 62.50 & 62.50 & 62.50 & 62.50 \\
Autoencoder embeddings (NEUROSKY--EPI only) & 58.33 & 58.33 & 62.50 & 58.33 \\
\bottomrule
\end{tabular}
\end{table}

This tabular comparison highlights two insights: (1) EEGNet embeddings provide stable performance across all algorithms, reflecting the benefits of transferring clinical knowledge, and (2) autoencoder embeddings, though more variable, can achieve comparable best-case performance while being self-contained and dataset-specific.

\subsection{Embedding Space Analysis}
To deepen interpretation of the results, we analyzed the embedding spaces using dimensionality reduction techniques. The EEGNet embeddings revealed clear patient-level groupings, with two main clusters corresponding to distinct EEG pattern characteristics. By contrast, the autoencoder embeddings exhibited a more fragmented structure: some patients formed tight subclusters, while others were more diffusely distributed in the embedding space. This fragmentation aligns with the clustering imbalance observed earlier.

These differences highlight two complementary strengths. EEGNet embeddings, grounded in clinical knowledge, create structured and transferable representations that align well with established seizure biomarkers. Autoencoder embeddings, trained only on local consumer-grade data, capture finer-grained but noisier individual variations. For practical deployment, these contrasts suggest a trade-off: EEGNet-based clustering may offer stability and interpretability, while autoencoder-based clustering could adapt more closely to context-specific populations, especially if expanded with more data.

From an HCI perspective, visualization of embedding spaces may also aid interpretability for clinicians and community health workers. If clusters can be visually inspected and understood, trust in the system is likely to improve, even if clustering accuracy is modest. In this sense, embedding analysis is not just a technical exercise but also a potential pathway toward more transparent and explainable decision support.

\subsection{Comparison with Traditional Patient Stratification}
Finally, it is important to situate these findings within the context of conventional patient stratification in epilepsy care. Clinical stratification typically integrates detailed seizure history, medication response, and multi-channel EEG analysis—approaches that can achieve high accuracy for specific subgroups but require specialist expertise and infrastructure. Our single-channel clustering pipeline, which achieved \textbf{62.50\% accuracy}, cannot replace these practices but demonstrates feasibility for \emph{automated, low-cost stratification} in settings where traditional resources are unavailable.

Notably, we did not identify prior work on unsupervised clustering of epilepsy patients using single-channel consumer EEG. Most existing studies focus on supervised seizure detection or multi-channel systems. Thus, our results substantially exceed random chance and represent, to our knowledge, a first step toward meaningful grouping with consumer devices. For resource-limited contexts, this performance suggests practical utility: clusters may provide an initial screening layer, flagging patients for specialist referral or supporting treatment monitoring when comprehensive evaluation is not feasible.

From a design perspective, the implications extend beyond accuracy. Balanced cluster sizes, explainable embeddings, and adaptability to local data are all characteristics that affect usability, fairness, and integration into health workflows. In this sense, our evaluation is not only a test of technical performance but also an exploration of how algorithmic clustering could function as a human-centered tool for equitable epilepsy care. These results, while promising, also raise broader questions about practical deployment, interpretability, and ethical use of clustering systems, topics that we explore in the following discussion.
\begin{figure}[ht]
\centering
\includegraphics[width=0.9\textwidth]{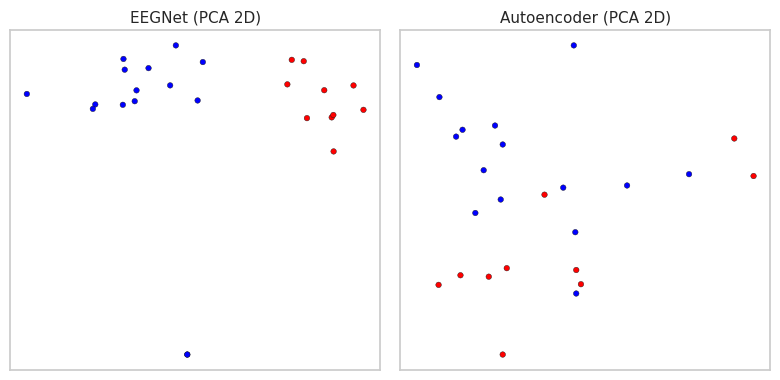}
\caption{Comparison of EEGNet approach vs Autoencoder}
\label{fig:pipeline-context}
\end{figure}

\section{Discussion}
Our findings demonstrate both the \textbf{promise and the challenges} of using consumer EEG devices for patient stratification in epilepsy care. Rather than positioning this as a fully mature solution, we critically analyze the practical, technical, and human-centered dimensions of such an approach. This includes examining the socio-technical implications of affordability, fairness across diverse patient populations, personalization potential, and the realities of clinical deployment. In doing so, we follow recent HCI scholarship that emphasizes not just technical feasibility but also responsible integration of health technologies into lived contexts.

\textbf{Affordability and Accessibility:} The NeuroSky MindWave headset used in this study costs on the order of \$100 USD—dramatically cheaper than multi-channel clinical EEG systems that can cost thousands. Such affordability creates an opening for EEG-based screening in communities that otherwise have no access to neurodiagnostic tools. From a health equity standpoint, this affordability is transformative: it aligns with HCI work on resource-constrained innovation, where inexpensive technologies can serve as “good enough” approximations for initial triage and monitoring. At the same time, \textbf{affordability does not guarantee adoption}. Devices must still contend with procurement policies, training requirements, and socio-cultural perceptions of legitimacy. Prior research on low-cost medical devices in low-resource settings shows that without trust-building and integration into local workflows, even inexpensive technologies can fail to scale. Thus, while our results demonstrate technical viability, rigorous deployment studies are needed to understand how cost interacts with usability, durability, and institutional acceptance.

\textbf{Diversity, Fairness, and Bias:} Our dataset includes patients from diverse educational, socioeconomic, and clinical backgrounds—an intentional step toward counteracting the homogeneity of existing EEG corpora. This diversity matters: epilepsy outcomes are shaped not only by neurophysiology but also by social determinants of health. Incorporating metadata such as seizure frequency change (15 patients with worsening, 10 without) enables stratification analyses that surface potential inequities. However, as fairness research in HCI and machine learning has shown, simply including diverse participants is insufficient; the downstream algorithms must also be stress-tested across subgroups to ensure they do not reproduce structural inequalities. For instance, clustering quality could differ systematically by age or comorbidity, which might exacerbate existing disparities in care. More rigorous fairness audits—similar to those conducted in algorithmic bias studies in healthcare—are required before deployment. Importantly, the open nature of NEUROSKY--EPI enables the broader community to interrogate these fairness issues, moving beyond single-lab evaluations.

\textbf{Clustering Refinement and Personalization:} One practical insight is the tension between \textbf{population-level models} and \textbf{individual-level personalization}. Our current system aggregates embeddings per patient to form group-level stratifications. While this yielded measurable clustering accuracy, HCI research on personalization suggests that static grouping may overlook important individual trajectories. For example, seizure likelihood is known to fluctuate with medication adherence, stress, and sleep—factors that evolve over time. A more rigorous system would therefore combine population-level embeddings with adaptive, longitudinal modeling that calibrates to a patient’s personal baseline. This personalization raises both opportunities and risks: it could make clustering more clinically useful, but it may also increase cognitive load for healthcare workers who must interpret dynamic changes. Future research should investigate hybrid models that balance interpretability with adaptivity, perhaps through clinician-in-the-loop approaches.

\textbf{Clinical Deployment and Workflow Integration:} Our results show that clustering can stratify patients above chance levels, but the key challenge is how such results would actually be used in practice. HCI research on decision-support systems emphasizes that clinical adoption hinges on \textbf{interpretability} and \textbf{actionability}. A clustering label alone is insufficient; clinicians must understand what distinguishes one group from another, and how those distinctions should inform treatment decisions. Without interpretability, the system risks being ignored or misused. Moreover, integrating a consumer EEG device into hospital or community workflows involves training, supervision, and compatibility with existing record systems. Prior experiences with mobile health technologies reveal that even promising prototypes often fail at this stage due to mismatches with clinicians’ routines or regulatory barriers. Thus, the rigor of deployment lies not only in accuracy but also in ensuring that clustering outputs are transparent, actionable, and aligned with institutional priorities.

\textbf{Rigor Beyond Accuracy:} Perhaps the most important lesson from our evaluation is that clustering accuracy alone does not determine impact. HCI and clinical AI research increasingly stress the need for \textbf{multi-dimensional evaluation}: usability studies with healthcare workers, qualitative assessments of interpretability, and ethical audits of bias and privacy risks. In our case, the 62.50\% accuracy is meaningful as a proof of concept, but further work must rigorously interrogate what such performance means for patient outcomes, how clinicians interpret probabilistic assignments, and how patients perceive being grouped by an algorithm. These questions extend beyond technical validation and require interdisciplinary collaboration between computer scientists, clinicians, ethicists, and patient advocates.

In summary, our discussion highlights that while low-cost EEG clustering offers exciting opportunities, a rigorous approach must attend to issues of affordability, fairness, personalization, and clinical integration. These socio-technical considerations are as critical as algorithmic performance if such systems are to advance from proof-of-concept to impactful, equitable healthcare tools.

\section{Limitations and Ethics}
\textbf{Current Limitations:} While the discussion highlights future opportunities, it is equally important to acknowledge the present limitations and ethical boundaries of our work. 
This work represents an early proof of concept for the stratification of patient populations based on consumer EEG, and its limitations must be carefully acknowledged. First, the clustering accuracy (62.50\%) is only moderately above chance. While this demonstrates feasibility, it is far from sufficient for any definitive clinical decision-making. The system provides \textit{probabilistic groupings} rather than diagnostic categories, and users must be reminded that such groupings are indicative rather than prescriptive. Second, the restricted spatial sampling of a single frontal electrode fundamentally limits the richness of patterns that can be captured; seizures often involve complex spatiotemporal dynamics across multiple brain regions that a one-channel device cannot observe. Third, our evaluation cohort (25 patients) is small and drawn from a single hospital context, limiting generalizability. Results may not extend to different epilepsy phenotypes, age distributions, or cultural settings. In addition, our stratification currently assumes a binary grouping based on seizure frequency change, while epilepsy manifests across a continuum of types, severities, and comorbidities. Future extensions could explore hierarchical or multi-class clustering to more faithfully reflect clinical complexity.

Another limitation lies in the interpretability of the clustering outputs. While clustering demonstrates above-chance accuracy, the system does not provide explanatory factors that distinguish clusters. Without interpretable features, clinicians may find it difficult to translate clusters into actionable treatment strategies. Moreover, the reliance on transfer learning from CHB-MIT introduces possible \textbf{domain shift}: EEGNet embeddings were pretrained on pediatric, multi-channel EEG, which may not fully align with adult and adolescent single-channel signals. Although our autoencoder approach partially mitigates this by training directly on NEUROSKY--EPI, the broader challenge of cross-dataset generalization remains unresolved.

\textbf{Ethical Considerations:} We adhered to strict ethical protocols throughout the study. All data were anonymized, replacing identifiers with codes and blurring faces in images. Public release of the dataset will only include de-identified features, with categorical or binned metadata to minimize re-identification risk. Institutional ethical approval was obtained, and participants (or guardians) gave informed consent with explicit clarification that the device was experimental and not intended for diagnosis. Communicating this boundary was essential to prevent undue clinical reliance or false hope.

Nonetheless, several ethical challenges remain. First, \textbf{algorithmic bias} is a real concern: EEGNet embeddings pretrained on pediatric CHB-MIT data may encode biases that disadvantage adults or underrepresented groups. While we attempted to address this through autoencoder embeddings, achieving fairness requires larger, more demographically diverse datasets. Second, informed consent in low-resource settings raises questions of comprehension and autonomy: participants must not only consent but also truly understand the experimental nature and limitations of such technologies. Third, potential harms include misclassification or stigmatization if clustering outputs are misinterpreted as definitive diagnoses. For example, a patient incorrectly grouped as “high risk” could face unnecessary anxiety or even changes in care pathways, while an under-classified patient might be denied specialist referral.

There are also governance concerns about \textbf{open datasets}. Once released, NEUROSKY--EPI could be repurposed by third parties for uses outside the scope of epilepsy care, raising questions about data stewardship and participant trust. Addressing this requires not only de-identification but also community-driven governance models that define acceptable use cases and accountability structures. Finally, the question of responsibility is critical: if clustering leads to misallocation of clinical resources, who bears liability—the clinicians, the developers, or the institutions deploying the system? The medical AI community continues to debate these issues, with growing consensus that such systems should be framed as \textbf{assistive decision-support tools} under human oversight, not autonomous decision-makers.

In conclusion, we remain optimistic that—with comprehensive validation, fairness audits, informed consent safeguards, and clear accountability frameworks—the benefits of accessible, consumer EEG-based stratification can outweigh the risks. However, achieving this balance requires interdisciplinary collaboration across clinical, technical, and ethical domains to ensure that innovations in affordable neurotechnology serve patients equitably and responsibly.

\section{Conclusion}
We presented \textbf{NEUROSKY--EPI}, a new open dataset and benchmark for single-channel EEG patient stratification, along with unsupervised clustering approaches that leverage deep neural embeddings for patient grouping. Our findings show that even a single $Fp1$ electrode from a consumer-grade headset can capture meaningful patterns for stratification, achieving 62.50\% clustering accuracy with both EEGNet embeddings (pretrained on CHB-MIT clinical data) and contextual autoencoder embeddings (trained directly on NEUROSKY--EPI). These results provide a proof-of-concept that \textbf{low-cost, wearable EEG devices} can contribute to actionable patient assessment, particularly in resource-limited settings where clinical infrastructure is scarce. The ability of both clinically-informed and dataset-specific embeddings to achieve comparable performance highlights the flexibility of our approach and its adaptability to different deployment contexts.

From a broader perspective, this work demonstrates the feasibility of \textbf{repurposing consumer technology for clinical decision support}. Just as smartphones have been adapted for health screening and monitoring, our results suggest that affordable EEG headsets can be reimagined as tools for stratifying patients when coupled with intelligent algorithms. For HCI and ubiquitous computing researchers, this opens new directions: designing devices that are not only technically accurate but also interpretable, trustworthy, and seamlessly integrated into clinical and community workflows. For healthcare stakeholders, it signals the possibility of building scalable, equitable infrastructures for neurological care using non-traditional hardware.

Looking forward, several avenues remain critical. First, we plan to \textbf{expand data collection} with larger and more diverse patient cohorts, longer recordings (including sleep), and multimodal data streams (e.g., accelerometry or heart rate) to improve robustness and generalizability. Second, we aim to develop \textbf{hierarchical stratification models} that move beyond binary clustering to capture the rich heterogeneity of epilepsy phenotypes. Third, we will conduct \textbf{real-world evaluations} in clinical and community settings to assess interpretability, workflow integration, and provider trust—key dimensions emphasized in HCI research. Finally, we will continue to prioritize \textbf{fairness, transparency, and ethics}, monitoring stratification performance across demographic groups and embedding safeguards for responsible data sharing and deployment.

In summary, our work takes an early but important step toward \textbf{inclusive epilepsy care}. We envision a future where anyone, anywhere, can access basic EEG-based stratification using inexpensive devices, with results contextualized and refined through human-centered design and clinical oversight. Achieving this vision will require interdisciplinary collaboration—bridging machine learning, HCI, clinical neurology, and ethics—to ensure that innovations in affordable neurotechnology are not only technically effective but also socially responsible and globally relevant. We believe NEUROSKY--EPI, together with the methods and reflections presented here, offers a foundation for such progress. 

\section*{Contribution Statement}
To summarize concretely, our contributions can be distilled into four key dimensions:

\begin{itemize}
    \item \textbf{Open Dataset:} We release \textit{NEUROSKY--EPI}, the first open-access epilepsy EEG dataset collected with a single-channel consumer headset in a low-resource context. The dataset (25 patients, 2,032 one-second windows) is accompanied by extensive medical and demographic metadata, enabling research not only on algorithmic performance but also on fairness, inclusivity, and the role of social determinants in epilepsy care.
    
    \item \textbf{Unsupervised Clustering Framework:} We propose a novel evaluation framework for patient stratification using affordable EEG hardware. By extracting deep neural embeddings from EEGNet (pretrained on clinical CHB-MIT data) and applying unsupervised clustering, we achieve 62.50\% accuracy across multiple clustering methods. This shifts the paradigm from event-level seizure detection toward population-level grouping, broadening how consumer EEG can be evaluated.
    
    \item \textbf{Dual Embedding Strategies:} We systematically compare two complementary embedding approaches: (1) clinically-informed EEGNet embeddings transferred from CHB-MIT, and (2) contextual autoencoder embeddings trained directly on NEUROSKY--EPI. Both strategies achieve comparable clustering performance (62.50\% and 58.33--62.50\%), demonstrating flexible pathways for deployment depending on data availability and resource constraints.
    
    \item \textbf{Human-Centered and Ethical Insights:} Beyond algorithmic performance, we foreground HCI considerations critical for real-world impact. We analyze clinical workflow integration, usability in low-resource settings, and patient diversity, while also reflecting on privacy, algorithmic bias, and appropriate use of clustering-based stratification tools. These insights aim to guide future interdisciplinary work at the intersection of machine learning, health technology, and human-centered design.
\end{itemize}


\end{document}